\documentclass{article}
\usepackage[english]{babel}
\usepackage{amsmath}
\usepackage{amssymb}
\usepackage{amsthm}
\usepackage{a4wide}
\usepackage[pdftex]{graphicx}
\usepackage{tabularx}
\usepackage{hyperref}
\usepackage{cite}

\newcommand{\qrbc}{\dot{\mathcal{Q}}_{RBC}}

\newcommand{\fah}{F\aa hr\ae us }

\title{Functional optimization of the arterial network.}

\author{Baptiste Moreau$^1$ and Benjamin Mauroy$^{1,2,\star}$\\
\\
\small{$\phantom{}^1$ Laboratoire MSC - UMR CNRS 7057,}\\
\small{Universit\'e Paris 7 / CNRS, Paris, France.}\\\\
\small{$\phantom{}^2$ Laboratoire J.A. Dieudonn\'e - UMR CNRS 7351,}\\
\small{Universit\'e de Nice-Sophia Antipolis, Nice, France.}\\\\
\small{$\phantom{}^{\star}$ Corresponding author, email: benjamin.mauroy@unice.fr.}}

\date{\today}

\begin{document}

\maketitle

{\bf Abstract.}

We build an evolutionary scenario that explains the selection of core physiological quantities of the arterial network of mammals. We propose that the arterial network evolved under the constraint of its function as an organ. To support this hypothesis, we focus on one of the main function of blood network: oxygen supply to the organs. We consider an idealized organ with a given oxygen need and we optimize blood network geometry and hematocrit with the constraint that it must fulfill the organ oxygen need. Our model accounts for the non-Newtonian behavior of blood, its maintenance cost and \fah effects. We show that the mean shear rates in the vessels follow a scaling law related to the multi-scale property of the tree network, and that this scaling law drives the behavior of the optimal hematocrit in the tree. Using experimental data, our model predicts an optimal hematocrit of  $0.43$ and an optimal ratio for blood vessels diameter decrease of about $0.79$, in agreement with physiology. Moreover our results show that pressure drops in the arterial network should be regulated in order for oxygen supply to remain optimal, suggesting that the amplitude of the arterial pressure drop may have co-evolved with oxygen needs.\\

\newpage

\section{Introduction}

By the process of evolution, biological structures are selected to be efficient both in their functions and in the costs their existence induces. Generally, identification of these functions and costs is very difficult and  relies on well chosen approximations. Organs have their own specificities since they are subparts of a whole organism and are intimately linked to other parts of the embedding organism \cite{weibel2}. Mammal vascular network is all the more specific, since it interacts with every part of the organism. The maintenance of a non optimized functional vascular network can be very costly. The first attempt to link the geometry of blood vessels with a cost was made by Murray in 1926 \cite{murray}. Blood was modeled with Poiseuille's laws and the cost was dissipation added to a metabolic rate of energy consumption in blood. This leads to Murray's law that states that minimal dissipation in a bifurcation is reached if the cube of the radius of the mother branch is equal to the sum of the cube of the radii of the daughter branches. This initial formulation has been extensively used and improved, and more general theories on networks have been developed, such as explanation for mammals or plants allometric laws \cite{west1,west2} or optimal design theory \cite{bejan}.

Murray's law has been found to be mostly valid for large blood vessels \cite{labarbera, zamir} but failed to describe properly microcirculation \cite{pries2}. Murray's law predicts that the wall shear stress remains constant in the whole circulation, but while this is quite true in large arteries, it is not in microcirculation \cite{zamir} where wall shear stress depends on blood pressure inside the vessels \cite{pries2}. Several studies have been focused on reformulating Murray's law to account for a wall shear stress depending on pressure in microcirculation, for example by introducing a cost for vasomotor control of blood flow by smooth muscles \cite{taber} or by accounting for non-newtonian blood rheology \cite{alarcon}. Unlike Murray's, most of these studies are however based on the hypothesis of a minimal cost at constant pressure drop between both ends of the network \cite{stark} with no control on the amount of blood flow. 

The hypothesis of a constant pressure drop between both ends of the network comes from the physiology of the cardio-vascular system: mammal blood pressures are regulated by metabolism \cite{comolet,stark} with very low variation between species \cite{west1}. Constraining pressure drop while searching the optimal properties of blood network hypotheses that there was a stage at which evolution could not alter pressure drop anymore. For example, a threshold for heart power output could have been met because of morphologic or metabolic limitations. However, pressure drop regulation could also be a "side-effect" of another optimization process. A process built on an {\it evolutive constraint} linked to blood network function would be closer to natural selection principles: individuals whose blood network was not able to fulfill the needs of the organs were eliminated by selection. Moreover, if such an optimization process predicts non physiological pressure drops, then it would support the hypothesis of the emergence of a pressure threshold during evolution.

The idea that structure evolves to optimize function arise from the symmorphosis theory \cite{taylor,weibelsym}. This theory postulates that "the state of structural design resulting from morphogenesis is regulated to match functional demand" (from \cite{weibelsym}). Typically, blood carries oxygen to the organs and for that specific function, functional demand would be the amount of oxygen flow needed by an organ. Thus blood network should be built in order to bring enough oxygen to the organs, but not too much to keep the unavoidable transport costs reasonable: a good candidate for the constraint would then be the quantity of oxygen driven by blood to the organ it feeds. This constraint mixes the capacity of the blood network to transport blood (geometry) with the capacity of blood to transport oxygen (red blood cells concentration) which in turn affects the capacity of the network to transport blood by altering blood viscosity. Thus our model has to take into account the rheology of blood, and the optimization should be performed not only on blood network geometry but also on red blood cells concentration in blood, i.e. hematocrit which represents the volumetric fraction of red blood cells in blood. Many studies have focused on the research of an optimal hematocrit \cite{birchard, crowell, schuler}, since an increased hematocrit enhances endurance performance and $\dot{V}O_{2max}$ \cite{ekblom, richardson, schuler, villafuerte}. It is however not clear if hematocrit should be artificially altered under pathological conditions \cite{ barone, czer, wan}. Actually, little is known on the consequences of an increased hematocrit on blood transport and cardiac output, and a fragile equilibrium in a patient status could be broken by therapeutic increase of hematocrit \cite{collins}. 

Finally, blood flows for rest and exercise regimes are different because of changes in oxygen needs. In some horses species, this goes along with hematocrit change: some horse breeds like Thoroughbreds and Quarter Horses have a highly variable hematocrit, about 0.35 at rest and up to 0.7 at exercise \cite{fedde}; Hematocrit is increased by the release of stocked red blood cells from the spleen \cite{fedde}. This suggests an adaptation of the configuration to the organ needs.

\section{Methods}

We will study an idealized organ that needs a constant oxygen flow to sustain its function. This organ is irrigated by a blood network that brings oxygen from an external source to the organ's capillaries. Our goal is to understand how the network geometry, the blood rheology and the blood fluid dynamics interact. We build an energy cost related to the organ oxygen supply and optimize this cost relatively to network geometry and blood hematocrit. The physical effects altering the energy cost are non linear and no trivial minimum can be found if red blood cells flow is constrained. The cost is hence minimized thanks to numerical simulations using Matlab (fonctions {\it fsolve} and {\it fminsearch}).

\subsection{Blood network properties}

\noindent{\bf{Blood network geometry.}} We consider an idealized "organ" whose oxygen supply is made with a blood network that connects an oxygen source to a capillary network. The blood network is modeled as a cascade of cylinders shaping a dichotomous tree. We neglect the role of the bifurcations. The oxygen source is located at the root of the tree and oxygen is distributed to the cells by a capillary network down each tree leaves. The capillary network, not accounted for in our model, constrains however the size of the leaves of the tree, connected to the capillaries. The number of bifurcations between a vessel and the root vessel of the tree defines its generation index. The tree root stands at generation $0$ and the tree leaves stand at generation $N$. We assume that the size of the vessels depends only on the tree root size and on its generation index (symmetric dichotomous tree \cite{mauroynat}). The largest arteries were not included because of the complex fluid dynamics occurring near the heart. Thus the tree root corresponds to a human vessel of about the sixth generation with a diameter $d_0=5 \ mm$ \cite{comolet}. The tree leaves correspond to large capillaries with diameters $d_N \sim 50 \ \mu m$.

\begin{figure}[h]
\centering 
\includegraphics[height=5cm]{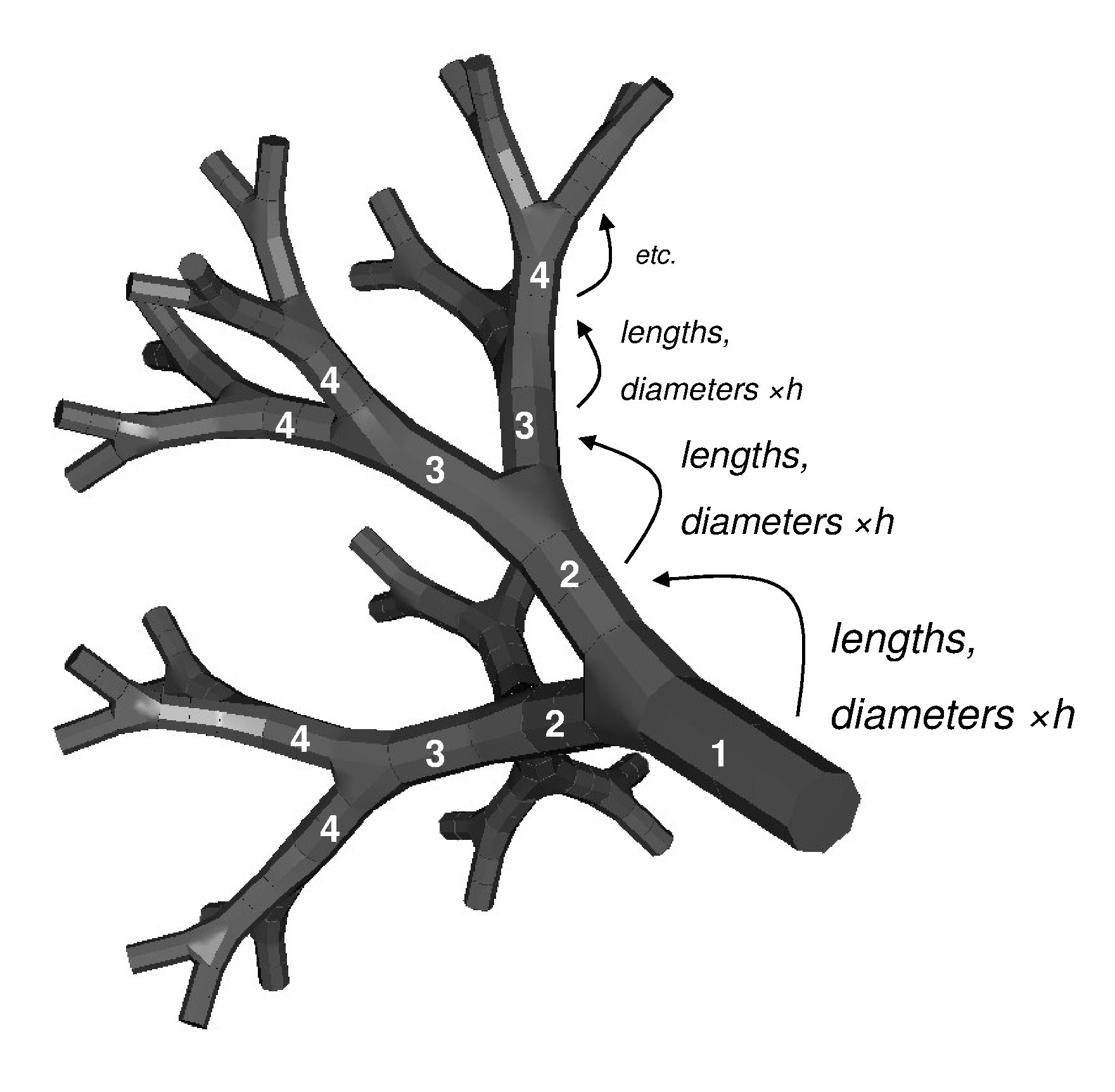}
\caption{Tree network structure: the tree is dichotomous and the vessels size decreases at each generation: their diameters and lengths are multiplied by the homothety factor $h<1$ after each bifurcation.}
\label{treegeom}
\end{figure}

We assume that vessels size and length decrease with the generation at a rate $h$, i.e. a vessel of generation index $i$ has a diameter $d_0 h^i$ and a length $l_0 h^i$. $h$ is called the vessel homothety ratio \cite{mauroynat}. In the following, the index $i$ will be used to refer to branches with generation index $i$. The number of generations $N$ that connect the root to the leaves depends on the homothety ratio $h$ and $N$ is computed by rounding $\log\left( d_N/d_0 \right) / \log(h)$. The root vessel length $l_0$ also depends on the homothety ratio $h$ to keep the length of a path from the tree root to one of the tree leaves constant and equal to $30 \ cm$ \cite{comolet}. The arterial network is a space-filling tree and its length does not reflect the length of the organ; moreover its total length does not influence the optimal geometry and hematocrit.

Oxygen is transported by blood in red blood cells, thus oxygen flow is proportional to red blood cells flow $\Lambda$. $\Lambda$ is equal to the mean volume of red blood cells going through a vessel of diameter $d_0$ \cite{comolet}: $\Lambda = 7.3 \ 10^{-7} m^3.s^{-1}$.\\

\begin{figure}[h!]
\centering 
\includegraphics[height=2cm]{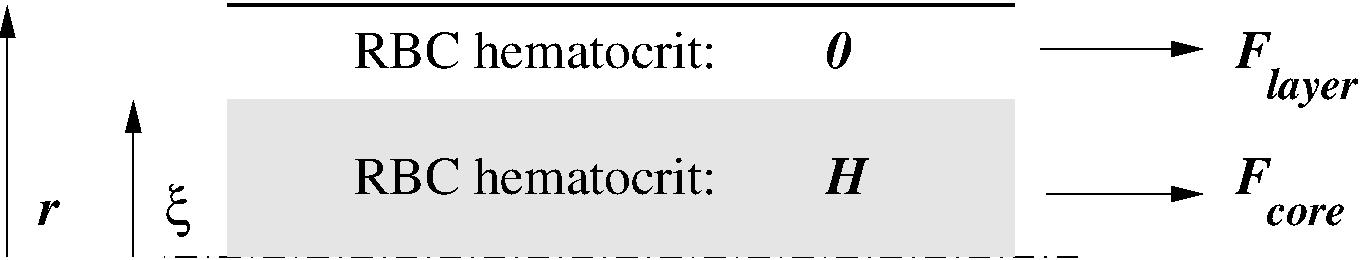}
\caption{Scheme of the blood flow inside an axi-symmetric vessel.}
\label{vesselscheme}
\end{figure}

\noindent{\bf Red blood cells distribution in a vessel.} We assume that red blood cells stand only in the core of the vessel where the hematocrit is constant and equal to $H$ (core hematocrit). The radius of the core is $\xi \leq r$. Plasma flows in the layer near the wall (its flow is $F_{layer}$). Plasma and red blood cells flow in the core (their flow is $F_{core}$), see figure \ref{vesselscheme}.

The flow of red blood cells in the core is $H \times F_{core}$. Since the total flow of red blood cells in the network $\qrbc$ is imposed, flow conservation implies for generation $i = 0, ..., N$:

$$
H_i \times F_{core,i} = \frac{\qrbc}{2^{i}}
$$

The red blood cells depletion near the vessel wall induces the F\r{a}hr\ae us effect that corresponds to a decrease of the mean hematocrit in the vessel relatively to the discharge hematocrit $H_D$ in large vessels. Since the blood that circulates into the vessel comes from larger vessels, by flow conservation \cite{fung, pries1} we can relate discharge hematocrit $H_D$ to core hematocrit $H$

$$
H_D = \frac{H F_{core}}{F_{core}+F_{layer}}
$$

The dependance of the mean hematocrit in a vessel, called tube hematocrit $H_T$, with tube radius and discharge hematocrit is well documented \cite{pries1}: If $d=2r$ is the diameter of the tube measured in microns, then

$$
H_T(H_D,d) = H_D \times \left( H_D+(1-H_D)\times (1+1.7 e^{-0.415 \times d} - 0.6 e^{-0.011 \times d}\right)
$$

Finally, tube hematocrit is related to core hematocrit by $H_T \pi r^2 = H \pi \xi^2$ and

$$
H_T=H \frac{\xi^2}{r^2}
$$

\subsection{Blood fluid mechanics}

\noindent{\bf Blood viscosity.} Blood is a non-Newtonian fluid that consists mostly in plasma and red blood cells. Plasma is a newtonian fluid about fifty percents more viscous than water. Blood behavior is mainly determined by red blood cells concentration in plasma and by local shear forces. Actually, red blood cells can aggregate if the surrounding forces are low and blood is rheo-fluidifying: an increase of shear stress in  decreases blood viscosity. Many models for blood viscosity were proposed \cite{Marcinkowska}; we chose Qu\'emada's model because it is well documented \cite{quemada} and accounts for both red blood cells volumetric fraction $H$ and shear rates $\dot{\gamma}$:

\begin{equation}
\mu(\dot{\gamma},H) = \mu_p \left( 1 - \frac{H}{H_{\infty}} \right)^{-2} \left( \frac{1+k}{\chi+k} \right)^2
\label{quemadalaw}
\end{equation}

The quantities $k$ and $\chi$ are defined by $k=\left( t_c |\dot{\gamma}| \right)^\frac12$, $\chi = \left( 1-\frac{H}{H_0} \right) / \left( 1-\frac{H}{H_\infty} \right)$; $\mu_P$ is the viscosity of plasma and $\mu_P=1.6 \ 10^{-3} \ Pa.s$. The quantities $t_c$, $H_0$ and $H_\infty$ are function of the red blood cells volumetric fraction $H$, their expressions have been estimated in \cite{cokelet}.

Blood viscosity varies strongly for shear rates ranging from $10^{-3} \ s^{-1}$ to $1 \ s^{-1}$ and reach a high viscosity plateau for shear rates smaller than $10^{-3} \ s^{-1}$ and a low viscosity plateau for shear rates larger than  $1 \ s^{-1}$. Viscosity increases with red blood cells volumetric fraction. Although hematocrit is an approximation of red blood cells volumetric fraction in blood, we will assume they are equal.\\

\begin{figure}[h!]
\centering
A 
\includegraphics[height=4.5cm]{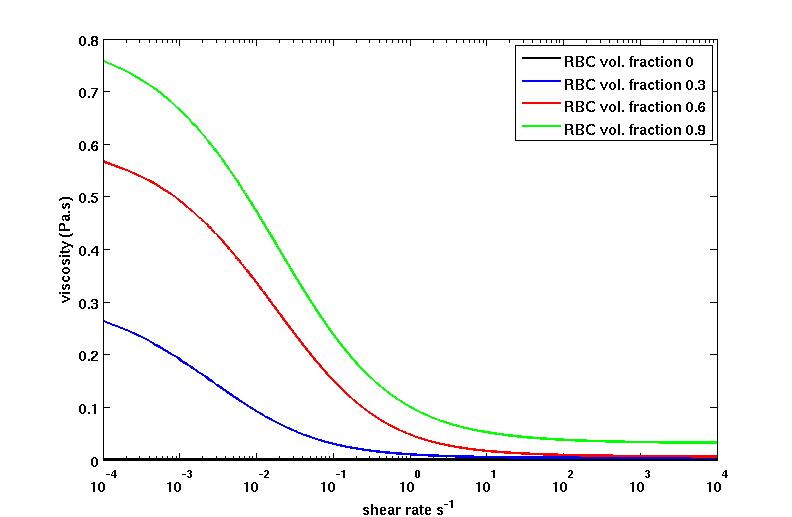}
B
\includegraphics[height=4.5cm]{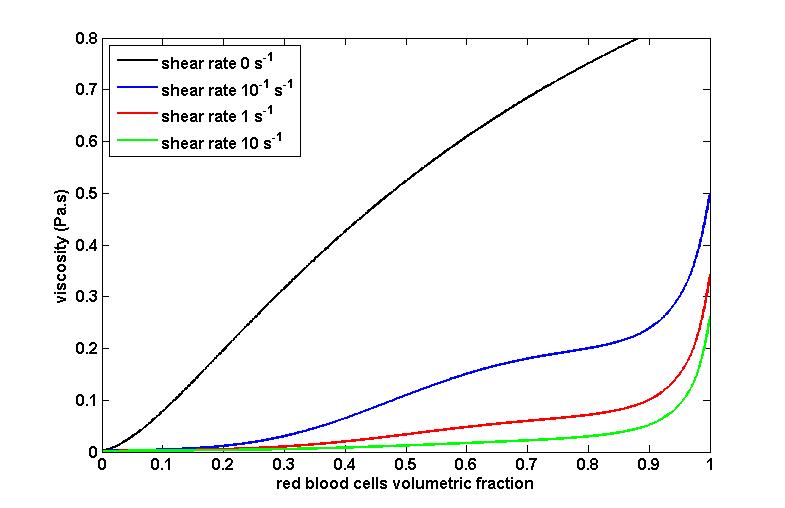}
\caption{Qu\'emada model of viscosity \cite{quemada, cokelet}. A: viscosity dependence on the shear rate for different values of the red blood cells volumetric fraction. B: viscosity dependence on the red blood cells volumetric fraction for different shear rates.}
\label{viscosityvs}
\end{figure}

\noindent{\bf Blood fluid equations.} The fluid is assumed axi-symmetric and fully developped in all the vessels. The viscosity $\mu(\dot{\gamma}(s),H(s))$ of the fluid at a radial position $s$ of a vessel depends on the local shear rate $\dot{\gamma}(s)$ and on the local red blood cell concentration $H(s)$: $H(s)=H$ in the core (if $0 \leq s \leq \xi$) and $H(s)=0$ in the layer (if $\xi < s \leq r$). The pressure drop per unit length is denoted $C$, it is assumed independent on the position. In this frame, blood fluid mechanics equations are:

$$
\mu\left(\dot{\gamma}(s),H(s)\right) \dot{\gamma}(s) = \frac{C s}{2}
$$

The viscosity near the wall is independent on the shear rate (no red blood cells) and $\mu_P=\mu\left(\dot{\gamma}(s),0\right)$. Finally, basic integral operations lead to core and layer flows

$$
F_{core} = -\int_0^\xi \dot{\gamma}(s) \frac{s^2}{2} ds + v(\xi) \frac{\xi^2}{2} \ \text{ and } \ F_{layer} = -\int_\xi^r \dot{\gamma}(s) \frac{s^2}{2} ds - v(\xi) \frac{\xi^2}{2}
$$
with
$$
v(\xi)=-\frac{C}{4 \mu_P} \left( r^2-\xi^2 \right)
$$

In the optimization process, the flow of red blood cells in the root vessel (generation $0$) is constrained: $\qrbc = H \ F_{core,0}$.

\subsection{Energy costs of oxygen transport}

Three major phenomena were identified as sources of energy consumption. First, blood viscous forces dissipate motion energy into heat energy in a rate $\mathcal{P}_D$; they must be counteracted by heart's pumping work. Secondly, red blood cells have a maintenance cost to keep efficient concentration gradients between their cytosol and blood plasma, and to maintain their membrane and hemoglobin. Third, red blood cells have a limited life span (about $120$ days \cite{weibel}) and new cells are produced continuously. In human, two millions new red blood cells are produced in red bone marrow each second as replacement for senescent ones \cite{weibel}. The cost can decompose into:

$$
\mathcal{C} = \underbrace{\sum_{\text{vessels}}\mathcal{P}_D(H,h)}_{\text{dissipation}} + \underbrace{\mathcal{P}_M(H,h)}_{\text{RBC maintenance}}+ \underbrace{\mathcal{P}_R(H,h)}_{\text{RBC renewing}}
$$

To store energy in ATP molecules, three main pathways exist, they are based on glucose degradation: Aerobic glycolysis is the most used and efficient, it is performed in mitochondria; Anaerobic glycolysis is less efficient but can be performed without mitochondria, it is the main source of energy for red blood cells; And finally glucose oxydation is marginally used by red blood cells.\\ 

\noindent{\bf Energy costs of blood circulation.} With our model's hypotheses, the dissipated energy during one second (power) in a vessel of radius $r$ and length $l$ writes
\begin{equation}
\mathcal{P}_D=2 \pi l \int_0^r \mu\left(H,\dot{\gamma}(s)\right) \dot{\gamma}(s)^2 s ds
\label{dissip}
\end{equation}
The amount of energy spent increases with fluid viscosity and with shear rates amplitude. However, blood is rheo-fluidifying (figure \ref{viscosityvs}A): if shear rate increases, then viscosity decreases. Thus, shear rate plays a complex role on dissipation.
The volumetric flow of red blood cells is constrained and depends on both the blood flow rate and the volumetric fraction of red blood cells (hematocrit). But one red blood cells flow rate can be reached with any  hematocrit value ($>0$) as soon as blood flow rate is adjusted, and these different possible configurations do not dissipate the same amount of viscous energy, since viscosity depends on hematocrit and on shear rates which in turn depend on blood flow rate.\\

\noindent{\bf Energy costs of red blood cells maintenance and replacement.} 
Red blood cells need energy to maintain efficient concentration gradients between their cytosol and blood plasma and to repair their membrane and hemoglobin. Because red blood cells do not contain mitochondria, this energy is produced by anaerobic glucose degradation ($90 \%$ using Embden-Meyerhoff pathway) and by glucose oxidation ($10 \%$ using hexose monophosphate shunt) \cite{baynes, hernberg} using blood glucose and oxygen. These nutrients are taken from the nutrients pool of metabolism, thus leading to a metabolic cost.

The most frequent pathway used by mammals to degrade glucose is the aerobic glycolysis. It is about eighteen times more efficient than anaerobic glycolysis with $36$ moles of ATP produced by aerobic degradation of one mole of glucose versus $2$ moles of ATP produced by its anaerobic degradation \cite{weibel}. Thus the energy withdrawn from blood by red blood cells in the form of glucose and oxygen is  potentially about eighteen times the energy actually used by red blood cells.

Typically, a red blood cell uses $5.56 \ 10^{-20} \ Mole.s^{-1}$ of glucose and $4.50 \ 10^{-19} \ Mole.s^{-1}$ of oxygen \cite{chow}. If those quantities of glucose and oxygen were consumed through the aerobic pathway, they would bring to the metabolism an energy of $\alpha_{M,G} = 1.59 \ 10^{-13} \ J.s^{-1}$ for the consumption of glucose and the corresponding amount of oxygen needed for its oxidation (six oxygen molecules for one glucose molecule), and of $\alpha_{M,O_2} = 5.56 \ 10^{-14} \ J.s^{-1}$ for the remaining oxygen \cite{weibel}. Finally, the maintenance cost is proportional to the number of red blood cells in the network, $N_{RBC}$, which corresponds to the ratio between the total volume of red blood cells in the network, $H \ \mathcal{V}(h)$, over the volume of one red blood cell, $v_{RBC}$:
$$
\mathcal{P}_M(H,h) = (\alpha_{M,G} + \alpha_{M,O_2}) \times \frac{H \ \mathcal{V}(h)}{v_{RBC}}
$$

Energy is also spent to replace senescent red blood cells. Red blood cells are produced in the red bone marrow from reticulocytes at a rate of about two millions cells per second \cite{weibel}. Relatively to the $2 \ 10^{13}$ red blood cells in a standard human body, this corresponds to a cell replacement rate of about $10^{-7}$ per red blood cell per second. The rate of oxygen consumption of a non mature red blood cell (progenitor) is about $1.80 \ 10^{-18} \ Mole.s^{-1}$ \cite{chow}. Progenitors become mature after four to seven days, thus we can estimate a mean oxygen consumption for the maturation of a new red blood cell to be about $1.08 \  10^{-12} \ Mole$ of oxygen, consequently the oxygen used per second per red blood cells in circulation for renewing is $1.08 \  10^{-19} \ Mole.s^{-1}$, i.e. an energy consumption of $\alpha_R = 5.20 \ 10^{-14} \ J.s^{-1}$. Finally, the red blood cells renewing cost is proportional to the number of red blood cells in the network:

$$
\mathcal{P}_R(H,h) = \alpha_{R} \times \frac{H \ \mathcal{V}(h)}{v_{RBC}}
$$

\section{Results/Discussion}

We first show that the optimal hematocrit in a single branch is directly related to the mean shear rate in the branch. Next, we show that mean shear rates in the tree follow a scaling law related to that of the tree. Finally, we show that this scaling law drives the behavior of the optimal hematocrit in the full tree.

\subsection{Optimal hematocrit in a single vessel} 

\begin{figure}[h!]
\centering 
A
\includegraphics[height=4.5cm]{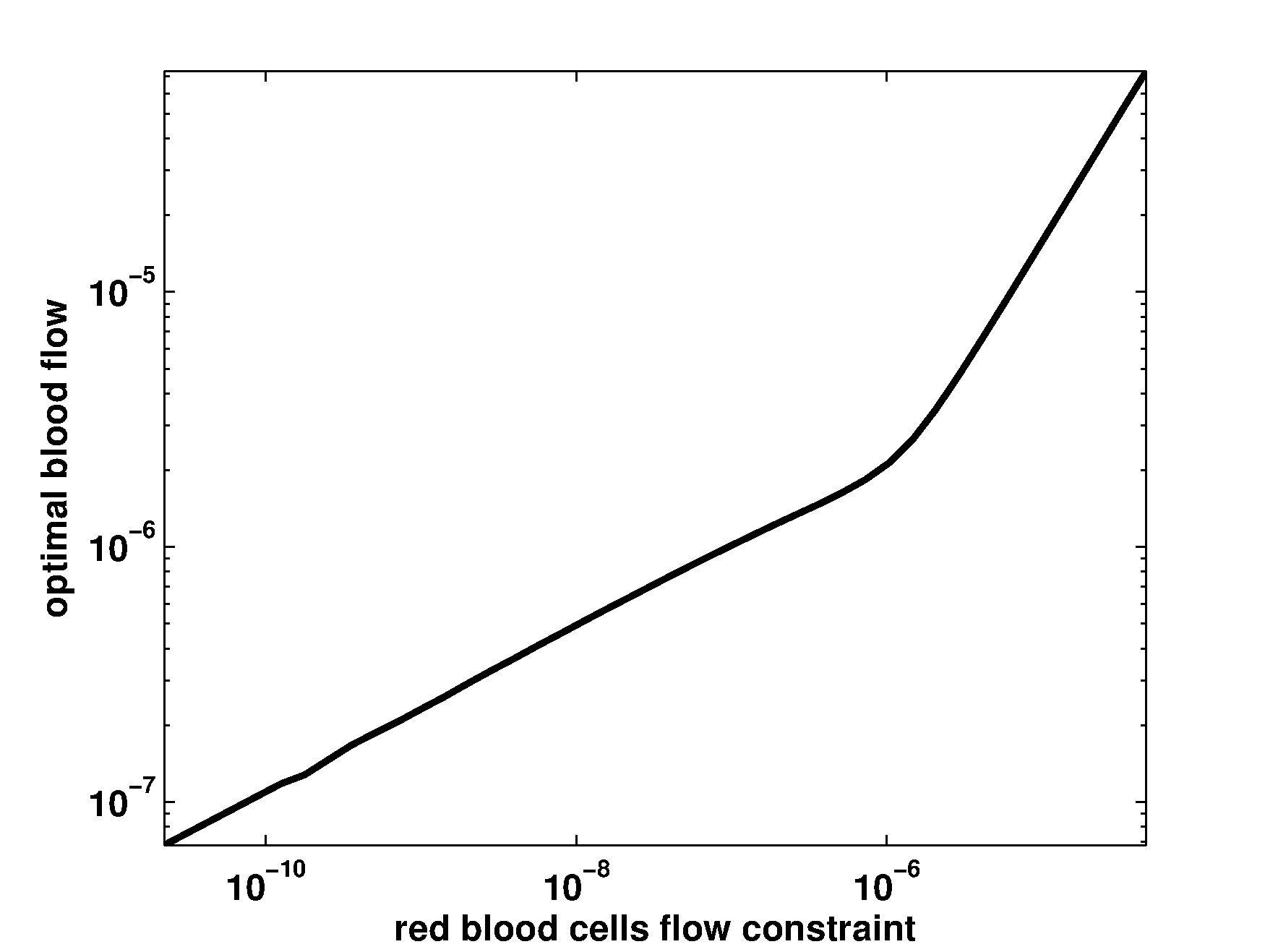}
B
\includegraphics[height=4.5cm]{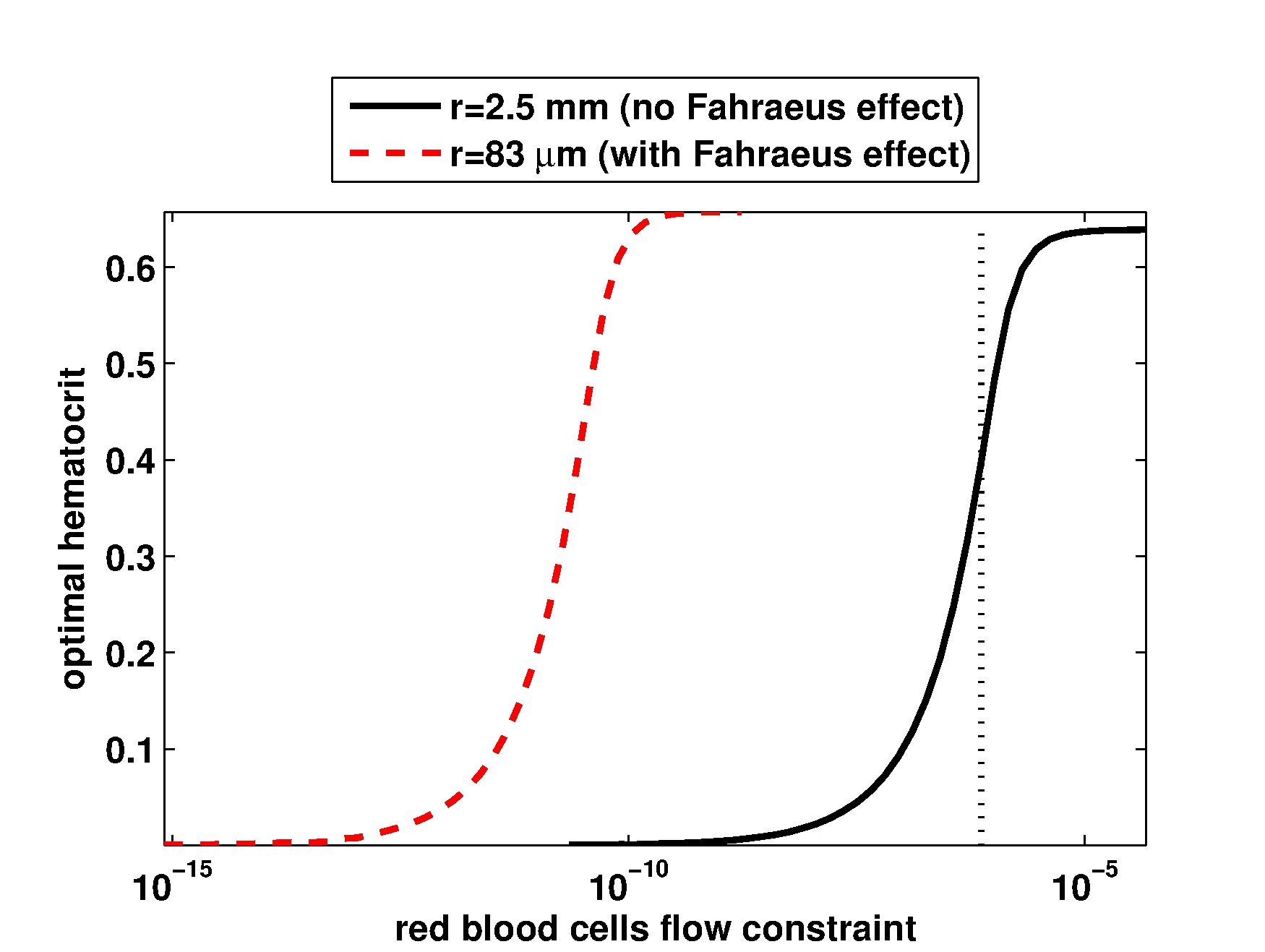}
\newline
C
\includegraphics[height=4.5cm]{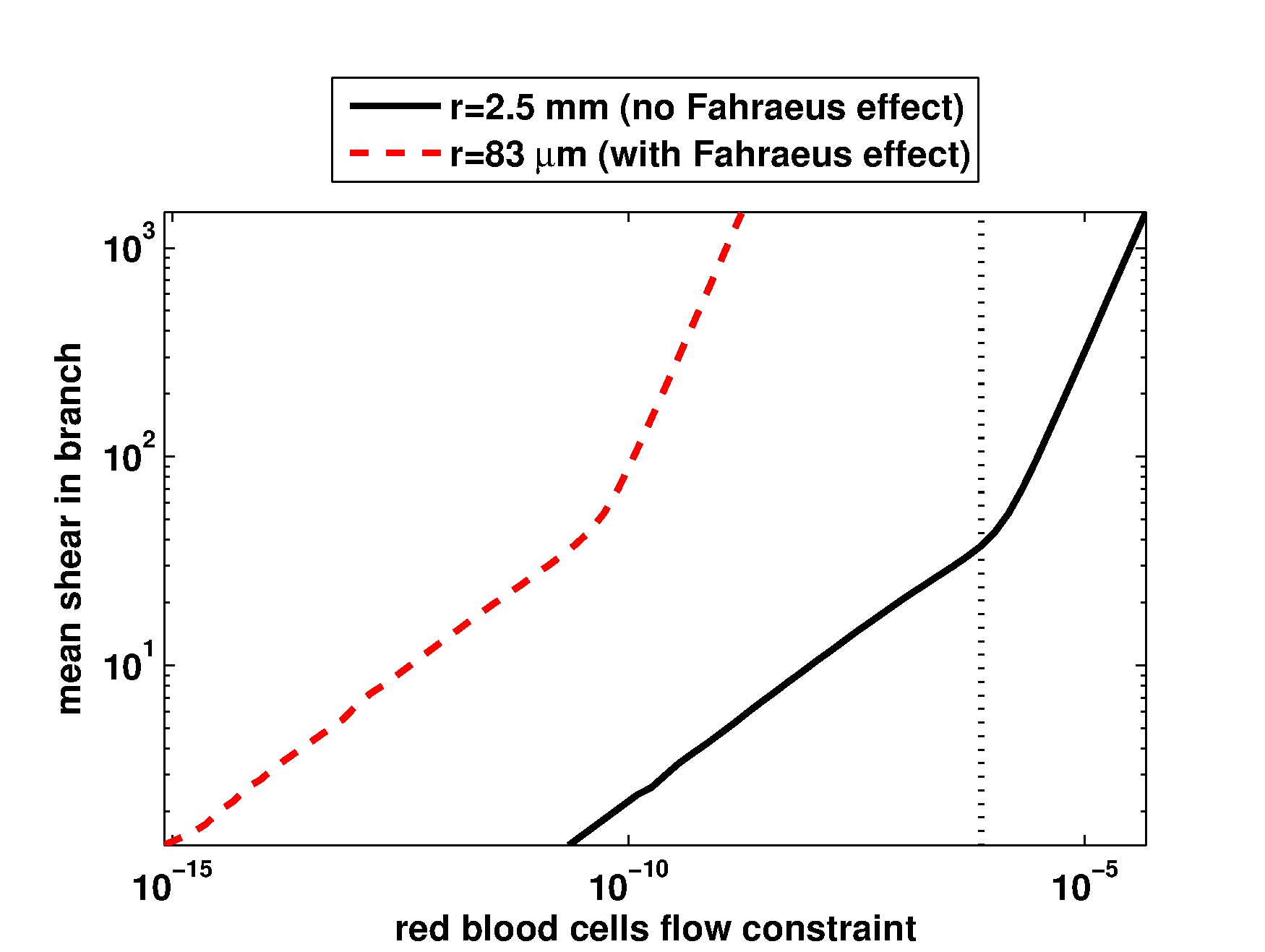}
D
\includegraphics[height=4.5cm]{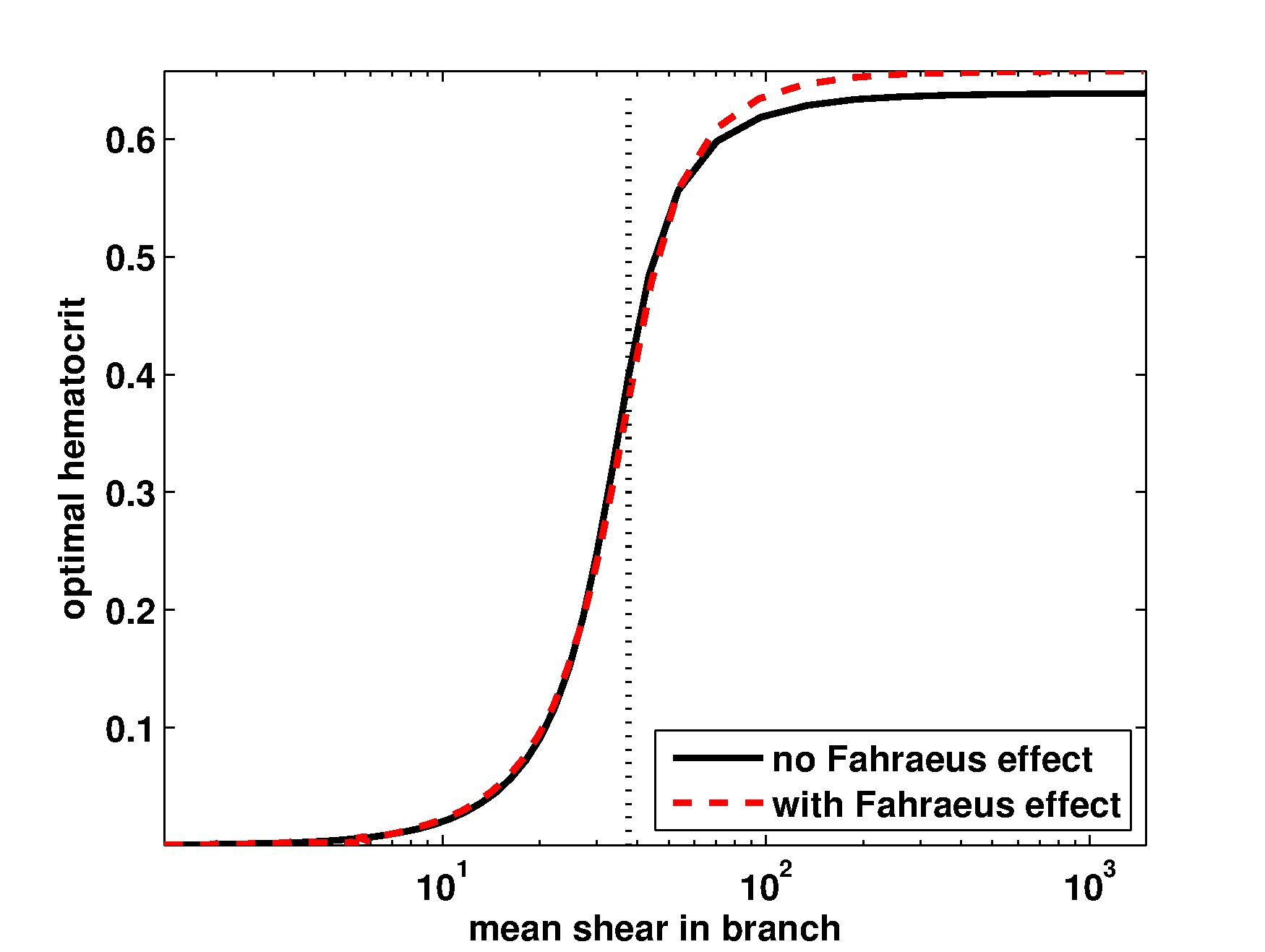}
\caption{A: Optimal blood flow versus red blood cells flow constraint, case of a branch whose radius is $2.5 \ mm$ (log log scale); B: Optimal hematocrit versus red blood cells flow constraint ($x$ axis in logarithmic scale); C: Optimal mean shear rate in the branch versus red blood cells flow constraint (log log scale); D: Optimal hematocrit versus mean branch shear rate with and without F\aa hr\ae us effect ($x$ axis in logarithmic scale). The vertical dotted line represents the mean shear rate estimated for the vessels in blood network. There is a one to one correspondence between the optimal mean shear rate in a branch and the optimal hematocrit. This correspondence does not depend on the branch diameter except if F\aa hr\ae us effect occurs, but then the dependence is slight.}
\label{shearplots}
\end{figure}

The optimal hematocrit in a single vessel will be referred to in the following as {\it vessel optimal hematocrit}. Because the volume of the branch does not change, the minimization of the energy costs is equivalent to the minimization of the dissipation in the branch. 
The optimization of the cost with red blood cells flow $\qrbc$ constrained, leads to values for both hematocrits and blood flow rates in the branch, which in turn brings optimal mean shear rates in the branch. Different configurations are plotted on figure \ref{shearplots}. 

Optimal states divide into two regimes separated with a sharp, but continuous transition, see figures \ref{shearplots}A and \ref{shearplots}B. The first regime corresponds to low red blood cells flow constraints: optimal states have low hematocrits ($H<0.1$) and high relative blood flow rates that compensate the low hematocrit. In this regime, most dissipation comes from the high flow in the branch since viscosity is almost minimal. When the constraint increases linearly, the optimal blood flow increases at a slower rate: $\qrbc \phantom{}^{0.32}$. When the optimal blood flow reaches the same order of magnitude than the red blood cells flow constraint, then the optimal hematocrit jumps to a value of about $0.65$. Then optimal blood flow vary almost proportionally to red blood cells constraints, like $\qrbc \phantom{}^{0.95}$.  In that second regime, dissipation is distributed between flow amplitude and median viscosity value. The viscosity in that second regime is about five to ten times larger than the viscosity in the first regime. 

Our simulations show that optimal mean shear rate in a branch is an increasing function of the red blood cells flow constraint, in the shape of a power law with two regimes, similar to that of the optimal blood flow since mean shear rate and blood flow in a branch are proportional, see figure \ref{shearplots}C.

Similarly, optimal hematocrit is an increasing function of the red blood cells flow constraint with two plateaus of low ($< 0.1$) and median hematocrit ($\sim 0.65$) for respectively low and high constraint values, with a sharp transition between plateaus, see figure \ref{shearplots}B. Consequently, the knowledge of the mean shear rate in a branch makes it possible to get back to the red blood cells flow constraint and then to retrieve the optimal hematocrit, see figures \ref{shearplots}B and \ref{shearplots}D. This is still true for small diameter branches subject to \fah effect, as shown by the red curves plotted on figure \ref{shearplots}. We computed and plotted on figure \ref{shearplots}B the dependency of optimal hematocrit on mean shear rate in a single branch with and without \fah effect. Although the dependence of both the optimal mean shear rate and the optimal hematocrit with red blood cells flow constraint changes with branch diameter, the dependence of the optimal hematocrit with the optimal mean shear rate is independent on the diameter when \fah effect does not occur (black curve on figure \ref{shearplots}D). If \fah effect occurs, then the dependence remains slight, as shown by the red curve on figure \ref{shearplots}D.

\begin{figure}[h!]
\centering 
\includegraphics[height=4.5cm]{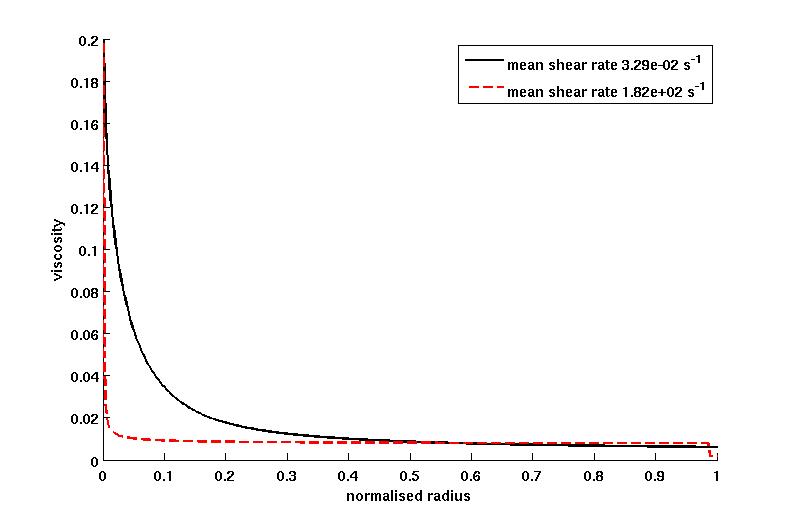}
\caption{Viscosity distribution along the radius of two branches with different mean shear rates. At shear rates larger than $10 \: s^{-1}$ (in the plateau of high viscosity on figure \ref{viscosityvs}A), the viscosity is almost constant, except near the center of the branch. At shear rates between the plateaus, the viscosity changes along the radius.}
\label{viscosity}
\end{figure}

The fact that the optimal hematocrit depends only on the mean shear rate in the vessel was to be expected since dissipation depends only on shear rate distribution in the branch and on hematocrit. Thus dissipation value is settled as soon as the distribution of shear rates in the branch and the value of hematocrit are known. But shear rates distribution in a vessel is fully determined by the mean shear rate in that vessel: there is a one to one correspondence between the value of the mean shear rate in the vessel and the profile of the shear rates in the vessel, as illustrated by figure \ref{viscosity}. Consequently, the knowledge of the mean shear rate in a vessel suffices to determine the vessel optimal hematocrit whatever the size of the branch, in the limit of \fah effect that slightly alters the dependence, see the red curve on figure \ref{shearplots}D. This property will be very helpful to understand the optimal hematocrit in a tree network, since we will show that the mean shear rates in the tree branches follow a scaling law.

\subsection{Optimal hematocrit in the tree network}

\noindent {\bf Behavior of shear rates in the tree structure.}
The tree structure is defined thanks to the homothety ratio $h$ that corresponds to the relative change in vessels diameter in a bifurcation, i.e. if $r_i$ is the radius of vessels of generation $i$, then the radii of vessels of generation $i+1$ are $r_{i+1} = h \times r_i$.

The regularity of the structure (fractal) leads to an interesting scaling law on how mean shear rates vary with generations as a function of the homothety factor $h$. Information on how the optimal hematocrit in the tree stands relatively to the optimal hematocrit in the branches of that tree can be deduced. The mean shear rate in a branch of generation $i$ is computed as the ratio between the mean blood velocity in the branch over the branch radius. Denoting $F_{T,i} = F_{core,i}+F_{layer,i}$ the total blood flow in that branch, and  $S_i = \pi r_i^2$ the surface of its circular cross section, then the mean shear rate in that branch is

$$
\left<\dot{\gamma}_i\right> = \frac{F_{T,i}/S_i}{r_i}=\frac{F_{T,i}}{\pi r_i^3}
$$

Since the tree is assumed dichotomous, the total blood flow in a branch of generation $i$ is twice the total blood flow in a branch of the next generation $i+1$. Then the mean shear rate in a branch of generation $i+1$ is

$$
\left<\dot{\gamma}_{i+1}\right> = \frac{F_{T,i+1}}{\pi r_{i+1}^3} = \frac{F_{T,i}}{\pi r_i^3} \frac{1}{2 h^3} = \left<\dot{\gamma}_{i}\right> \frac{1}{2 h^3}
$$

Consequently, the mean shear rate follows a scaling law. This scaling law depends only on the homothetic ratio $h$. Depending on the position of the factor $\frac{1}{2h^3}$ relatively to $1$, the mean shear rate has different behaviors:

\begin{itemize}
\item if $h>\left(\frac{1}{2}\right)^{1/3}$, then the mean shear rate decreases along the generations, consequently blood viscosity tends to increase along the generations.
\item if $h=\left(\frac{1}{2}\right)^{1/3}$, then the mean shear rate remains constant along the generations and so for blood viscosity.
\item if $h<\left(\frac{1}{2}\right)^{1/3}$, then the mean shear rate increases along the generations, consequently blood viscosity tends to decrease along the generations.
\end{itemize}

Blood viscosity depends on shear rate in a non linear way, with a plateau of high viscosities at low shear rates and a plateau of lower viscosities for high shear rates. Between the two plateaus, for medium shear rates, blood viscosity varies quickly, as shown on figure \ref{viscosityvs}A. When the shear rate decreases or increases along the generations of the tree, then viscosity will vary more strongly if shear rate variation makes the viscosity go through the strongest slope. Consequently, the amplitude of viscosity variation depends on the initial mean shear rate $\left<\dot{\gamma}_0\right>$ in the root branch of the tree: a high shear rate in the root branch can induce notable viscosity variations throughout the tree only if shear rate is decreasing enough along generations; similarly, a low shear rate in the root branch can induce notable viscosity variations throughout the tree only if the shear rate increases enough along the generations. In any other case, the viscosity is stalled, either in one of the two plateaus or because mean shear rate does not vary much if $h \sim \left(\frac{1}{2}\right)^{1/3}$. These results are summarized in table \ref{tableviscosity}.\\

\begin{table}[h!]
\begin{center}
\begin{tabular}{|c|c|c|c|}
\hline
 & $h < \left(\frac{1}{2}\right)^\frac13$ & $h=\left(\frac12\right)^\frac13$ & $h>\left(\frac12\right)^\frac13$\\
\hline
\begin{tabular}{c} 
low $\left<\dot{\gamma}_0\right>$\\
(high viscosity in root) 
\end{tabular} 
& decrease of viscosity & \begin{tabular}{c} high constant\\ viscosity \end{tabular} & \begin{tabular}{c} high viscosity,\\ small negative\\ variations \end{tabular}\\
\hline
\begin{tabular}{c} 
high $\left<\dot{\gamma}_0\right>$\\ 
(low viscosity in root) 
\end{tabular}
& \begin{tabular}{c} low viscosity,\\ small positive\\ variations \end{tabular} & \begin{tabular}{c} low constant\\ viscosity \end{tabular} & increase of viscosity\\
\hline
\end{tabular}
\caption{Behavior of viscosity along the generations as a function of the homothety ratio $h$ and the initial mean shear rate $\left<\dot{\gamma}_0\right>$.}
\label{tableviscosity}
\end{center}
\end{table}

\noindent{\bf Optimal hematocrit in the tree.}
Dissipation depends on two quantities: viscosity and shear rate. If the shear rate increases along the generations, as for the case $h<\left(\frac{1}{2}\right)^{1/3}$, then dissipation induced by shear rates increases, however the dissipation induced by viscosity decreases since viscosity decreases when shear rate increases. 

Red blood cells maintenance and renewing are proportional to network volume and network volume increases with $h$. For $h$ smaller than $\left(1/2\right)^{1/3}$, most of the volume of the network stands in the first generations; For $h$ larger than $\left(1/2\right)^{1/3}$ most of the volume stands in the last generations. Consequently, for $h$ smaller than $\left(1/2\right)^{1/3}$, energy cost arises mostly from dissipation induced by high shear rates, since viscosity is low and red blood cells maintenance cost is minimal (low network volume).  For $h$ larger than $\left(1/2\right)^{1/3}$, energy cost arises mostly from dissipation induced by high viscosity and from red blood cells maintenance cost that is maximal (high network volume).

\begin{figure}[h!]
\centering 
A
\includegraphics[height=5cm]{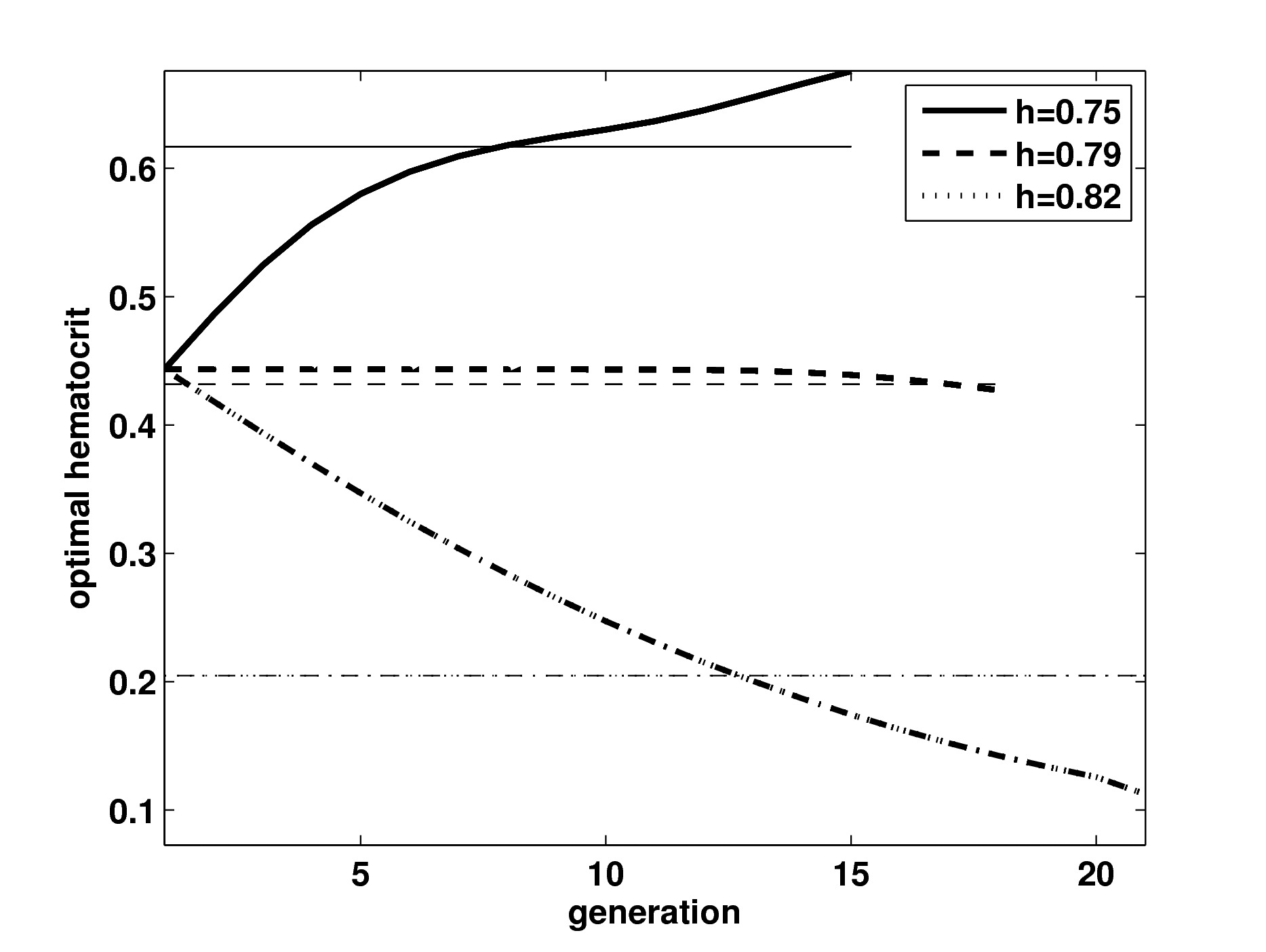}
B
\includegraphics[height=5cm]{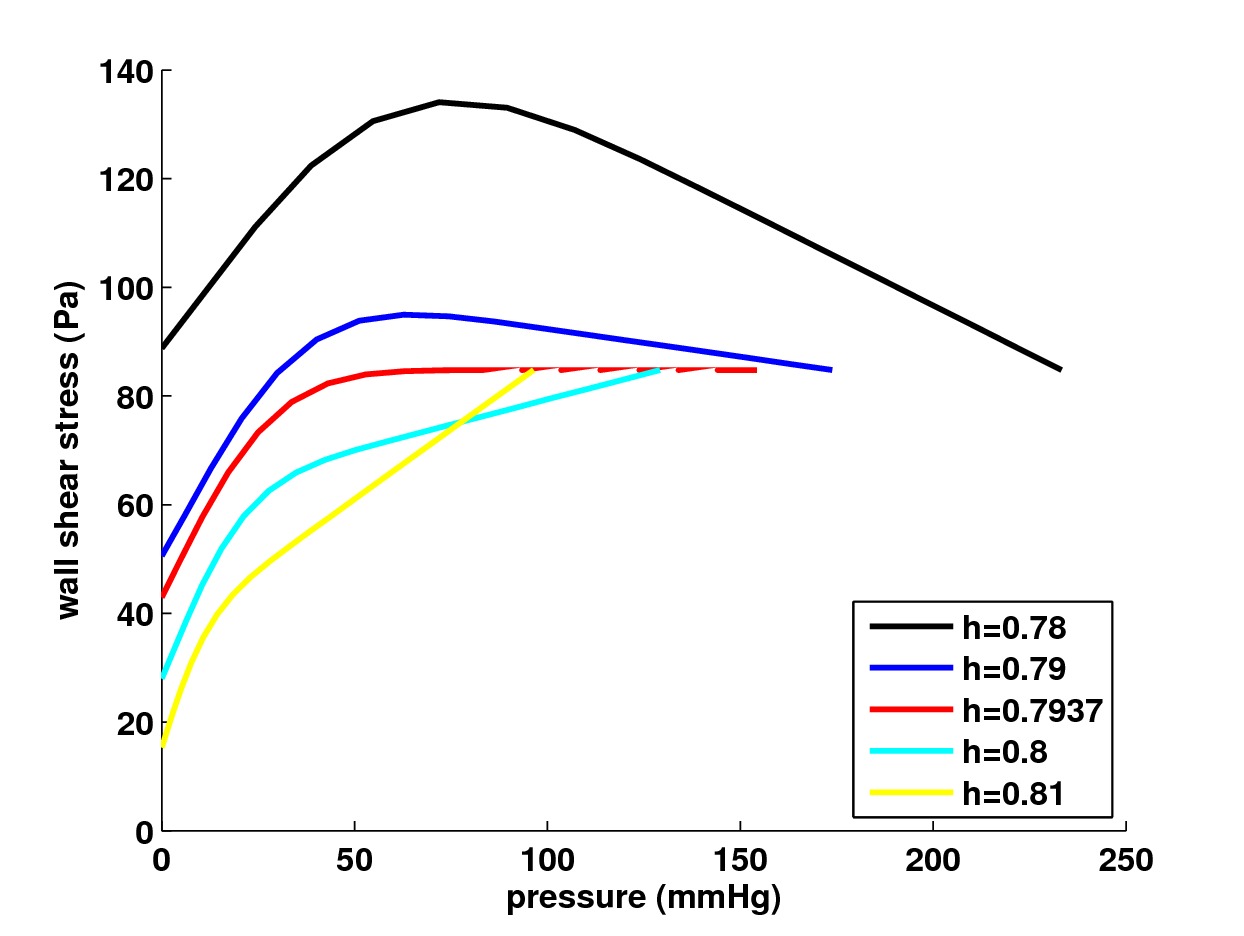}
\caption{A: Optimal hematocrit computed in each branch (thick lines) and compared with the optimal hematocrit for the whole tree (horizontal thin lines). Three values of $h$ are plotted: $h=0.75$, $h=0.79$ and $h=0.82$. B: Relationship between wall shear stress and pressures in the vessels for different values of the homothety ratio $h$.}
\label{optHBr}
\end{figure}

This behavior is illustrated on figure \ref{optHBr}A that shows the optimal hematocrit in each branch of the tree (thick lines) and the global optimal hematocrit for the tree (thin horizontal lines) for three different geometries: $h=0.75$, $h=0.79 \sim \left(\frac12\right)^\frac13$ and $h=0.82$. The red blood cells flow needed by the idealized organ fits the typical physiological amount that goes through a mammal arterial tree of the size of our model $\qrbc = 7.3 \ 10^{-7} \ m^3.s^{-1}$ \cite{comolet, west1}.

As expected, mean shear rates in the branches increase along the generations for $h=0.75$ and thus, as stated in the previous section (figure \ref{shearplots}D), the vessel optimal hematocrit increases with the generation. In that case, the tree optimal hematocrit is about $0.62$. For $h=0.79$, mean shear rate in the tree is almost constant along the generations and so are the vessel optimal hematocrits. The optimal hematocrit for the tree is then about $0.42$. Finally, if $h=0.82$, mean shear rate in the tree is decreasing along the generation and so are the vessel optimal hematocrits in the branches. In that last case, the optimal hematocrit in the tree is about $0.205$.

Our model predicts that wall shear stress is not constant along the generations, in agreement with the observations \cite{pries2}. The pressure inside the vessels is decreasing along the generations, the dependance of wall shear stress on pressure inside the vessel is plotted on figure \ref{optHBr}B. Wall shear stress is always an increasing function of pressure when pressure is small (small vessels). For larger pressures (large vessels), wall shear stress is increasing if $h>(1/2)^{1/3}$, is constant if $h \sim (1/2)^{1/3}$ and is decreasing if $h<(1/2)^{1/3}$. The case that best fits the physiology is $h \sim (1/2)^{1/3}$ \cite{alarcon, pries2, taber}.

\subsection{Optimal irrigation of the organ}

In this section, both optimal hematocrit value and homothety ratio $h$ are optimized for them to minimize the total energy needed to sustain the network while feeding the idealized organ. We still consider the same typical flow of red blood cells in the root of the tree: $\qrbc = 7.3 \ 10^{-7} \ m^3.s^{-1}$ \cite{comolet}. 

\begin{figure}[h!]
\centering
A
\includegraphics[height=5.5cm]{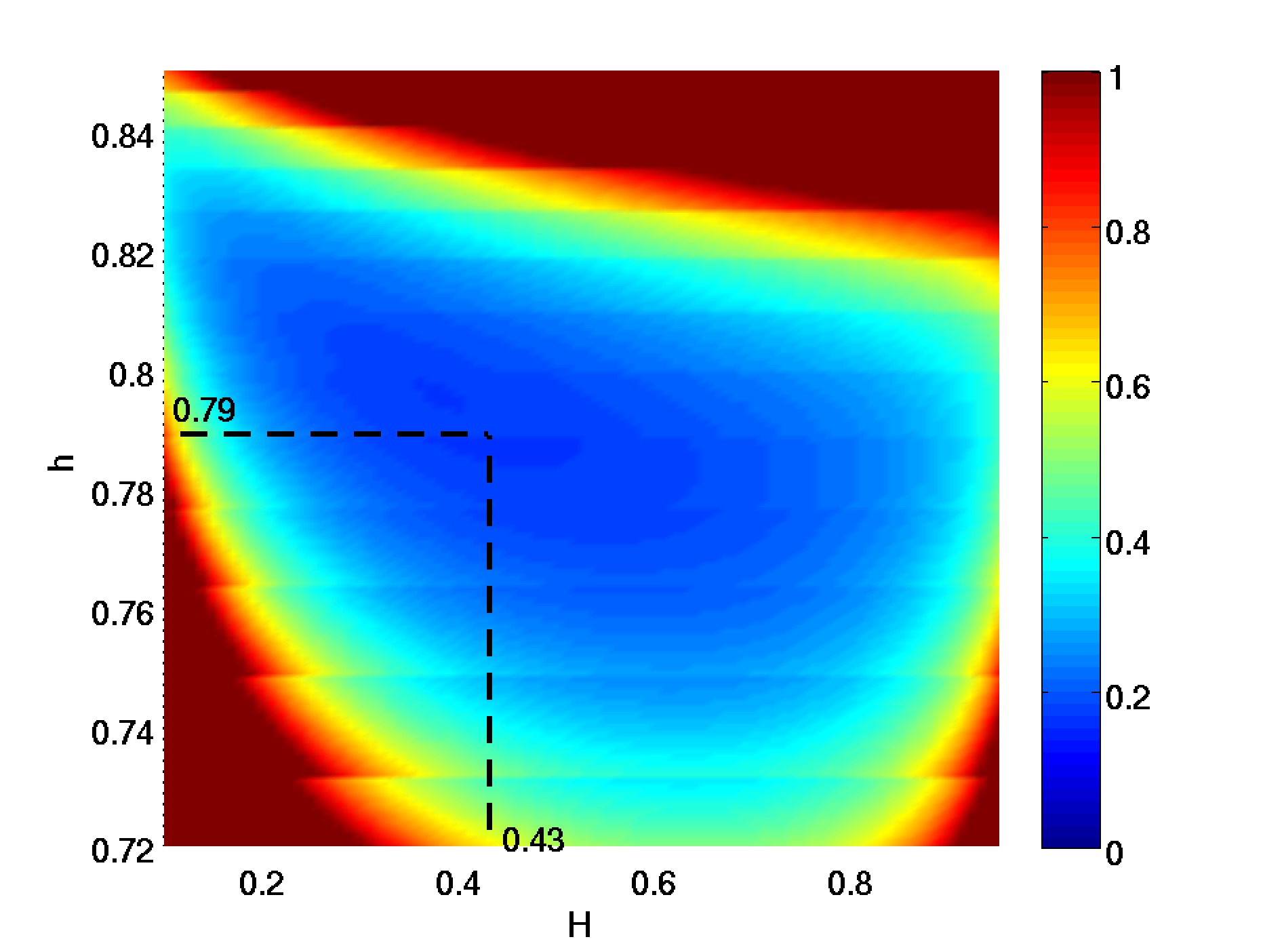}
B
\includegraphics[height=5cm]{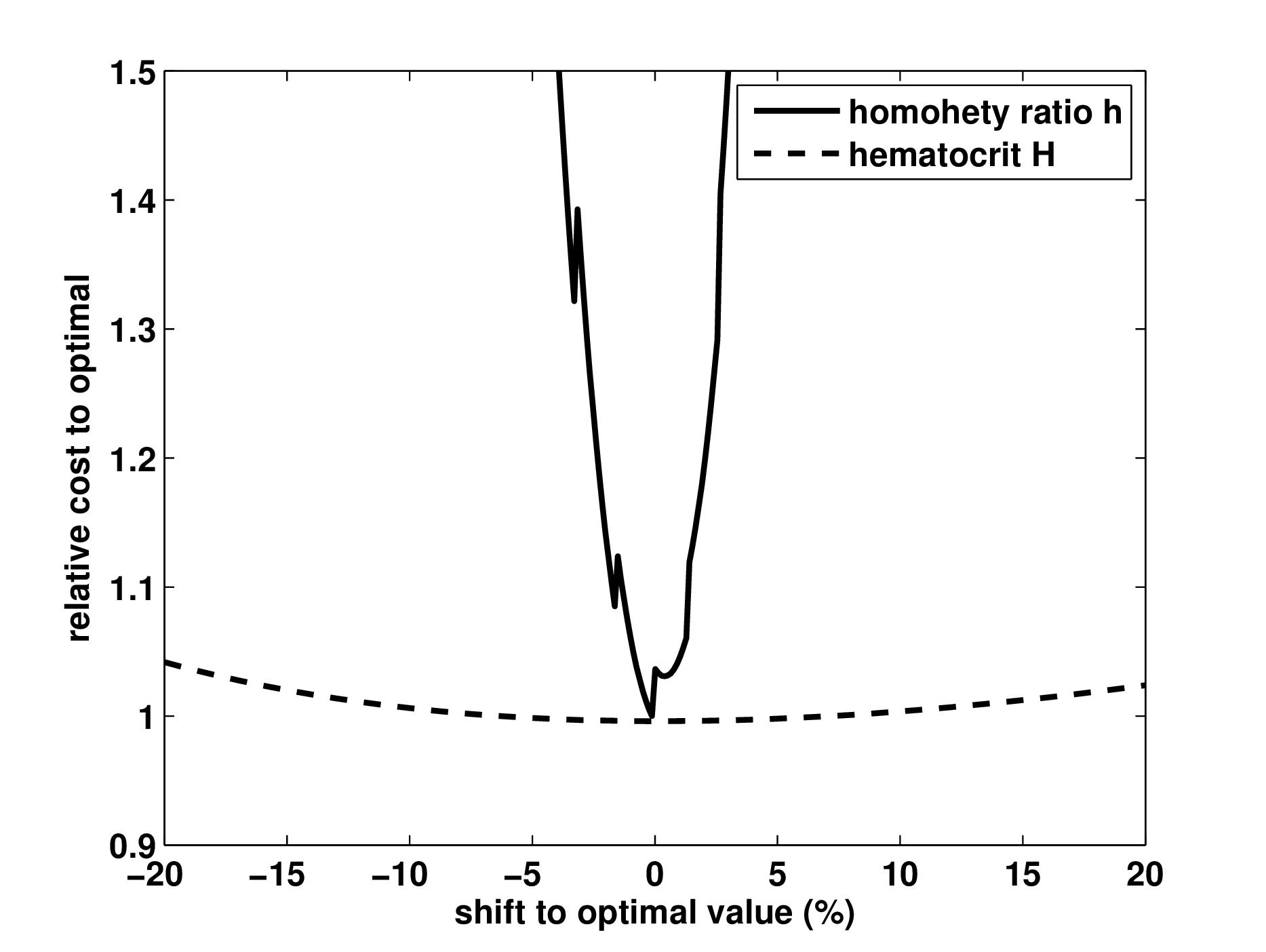}
\caption{A: Energy cost to irrigate the idealized organ (in $J.s^{-1}$) versus hematocrit $H$ and homothety ratio $h$. The minimum is reached for $h=0.79$ and $H=0.43$. B: Sensitivity of normalized energy cost to $h$ and $H$ variations. Energy cost has low sensitivity to hematocrit variation in the physiological rangel but has high sensitivity to geometric variations of the network.}
\label{energylow}
\end{figure}

Our model predicts that the organ irrigation is optimal when the homothety factor is $h=0.79$ and the hematocrit is $H=0.43$. These results fit closely the physiology, indicating that the portion of the tree modeled in this work may have played an important role in hematocrit value selection. The energy landscape as a function of hematocrit $H$ and homothety ratio $h$ is plotted on figure \ref{energylow}A. The landscape has discontinuities (seen as horizontal color breaks on the figure \ref{energylow}A) that correspond to changes in total generations number (see geometry section). Near the optimal value, located at the intersection of the dashed lines on figure \ref{energylow}A, energy has a flat variation along hematocrit axis as shown by the dashed curve on figure \ref{energylow}B. This indicates that the system is robust relatively to hematocrit change: if hematocrit varies around the optimum and stays in the physiological range (say from about $0.3$ to about $0.5$), then energetic cost does not change much, although oxygen transport is either reduced or increased, proportionally to hematocrit. Low hematocrits often lead to anemia and body response is to increase oxygen transport capacity of blood by stimulating erythropoiesis \cite{storz}. High hematocrits, up to $0.6$ or even $0.7$, can also be encountered, either in pathologies (polycythemia), in medical care or in doping (erythropoietin or auto-transfusion). 

Our model can bring quantitative data about the consequence of a strong increase of hematocrit. It shows that an increase in hematocrit from the optimal value $0.43$ to $0.60$ ($+40 \ \%$) increases the energetic cost for the organ irrigation of $46 \ \%$. Dissipation in the network increases of about $51 \ \%$ and red blood cells maintenance of about $40 \ \%$. If no other effects are limiting the transfer of oxygen to the cells, the gain of oxygen would be about $40 \ \%$. As expected, our model predicts that oxygen availability is increased, but heart would also be submitted to a much higher blood resistance, increasing tiring and heart failure risks.

Energetic cost is very sensitive to homothetic ratio $h$, see figure \ref{energylow}B. This was to be expected since the size of the vessels is very sensitive to the parameter $h$. Diameters, and lengths in a less extent, are directly correlated to both blood dissipation and total volume \cite{mauroynat}. Actually, the sensitivity is due to the multiplicative process induced by the parameter $h$: the diameter and length of a branch at generation $i$ are proportional to $h^i$ \cite{mauroynat,bokov}.\\

\begin{figure}[h!]
\centering
A
\includegraphics[height=5cm]{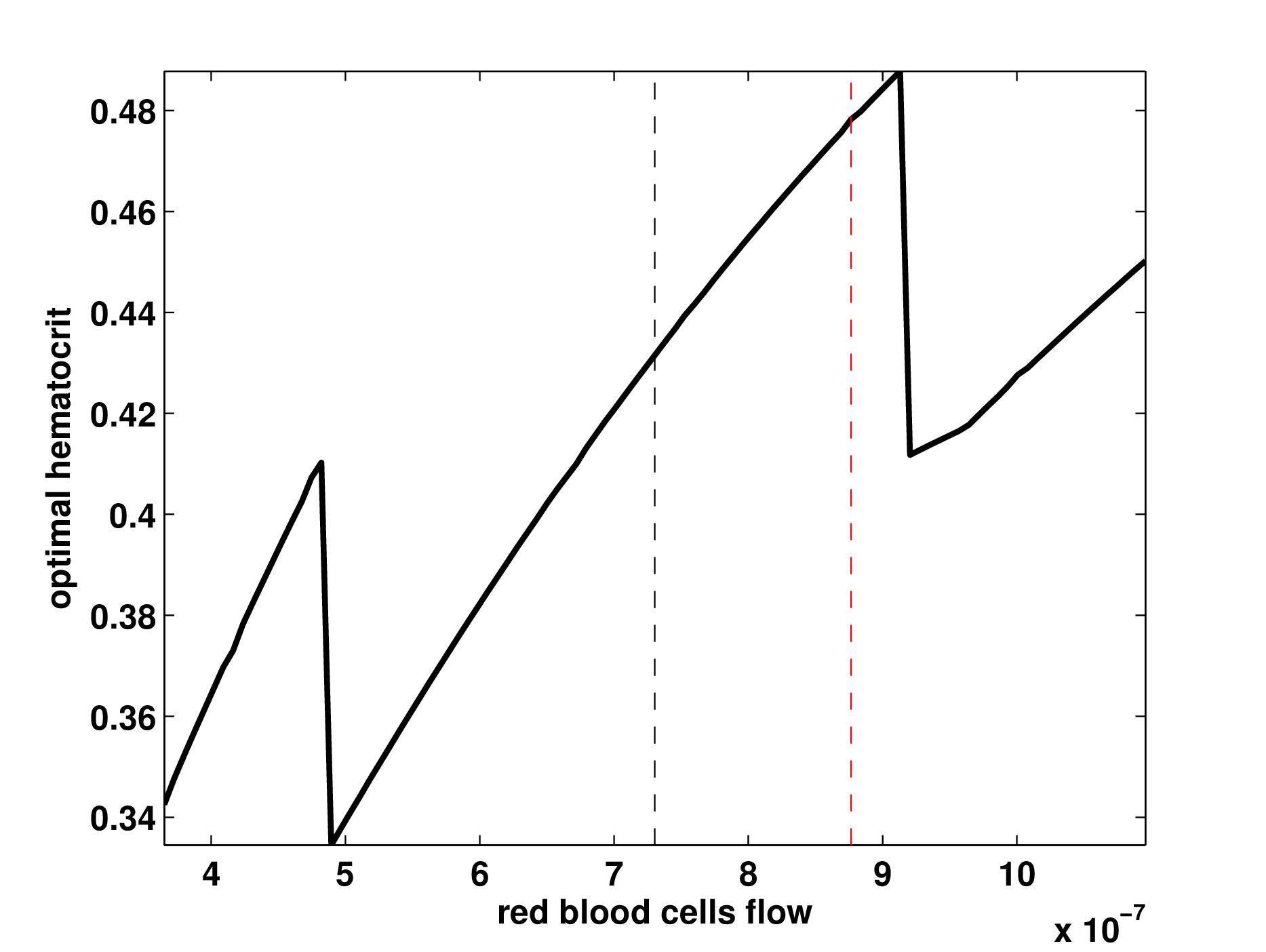}
B
\includegraphics[height=5cm]{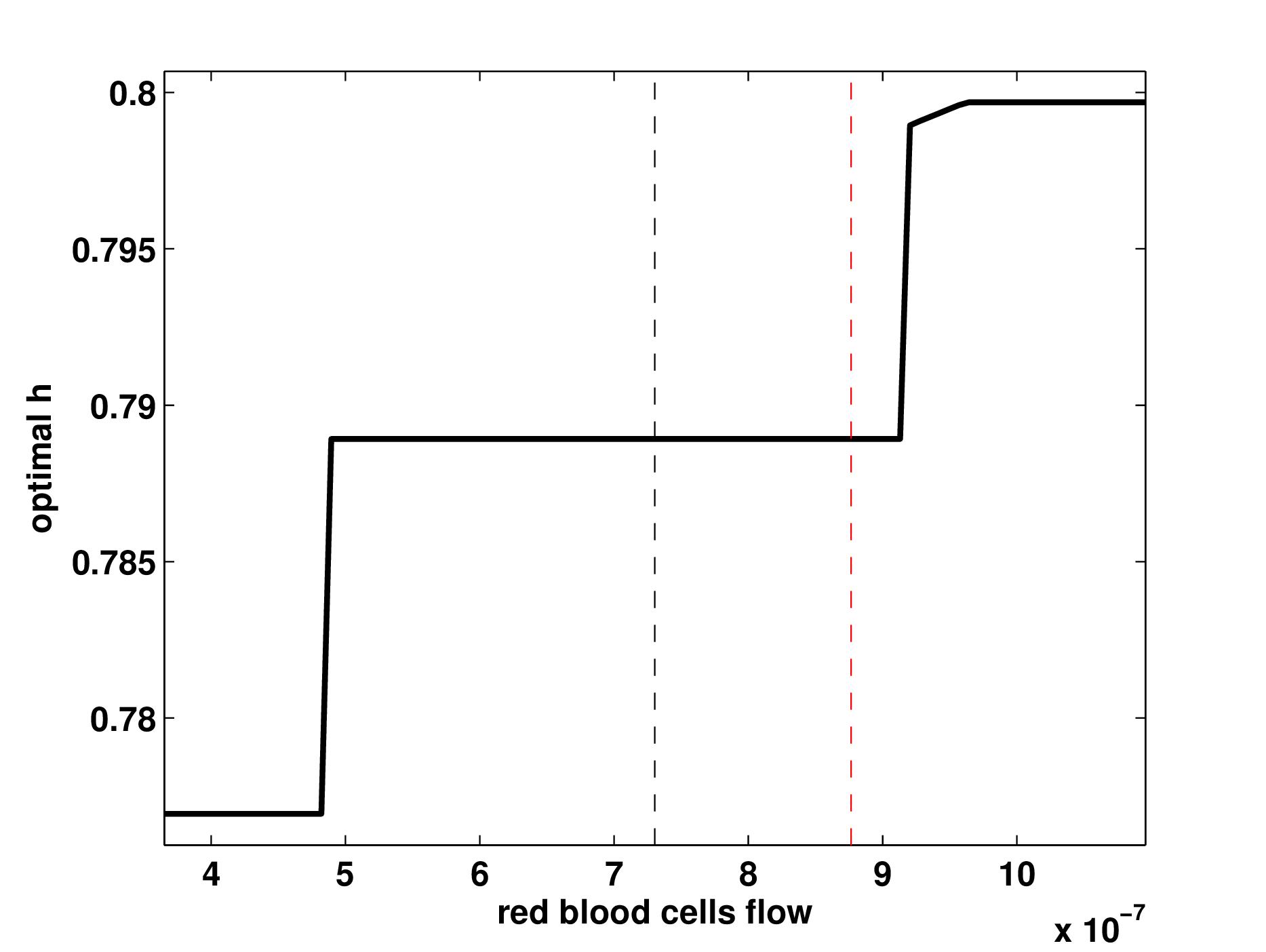}
\newline
C
\includegraphics[height=5cm]{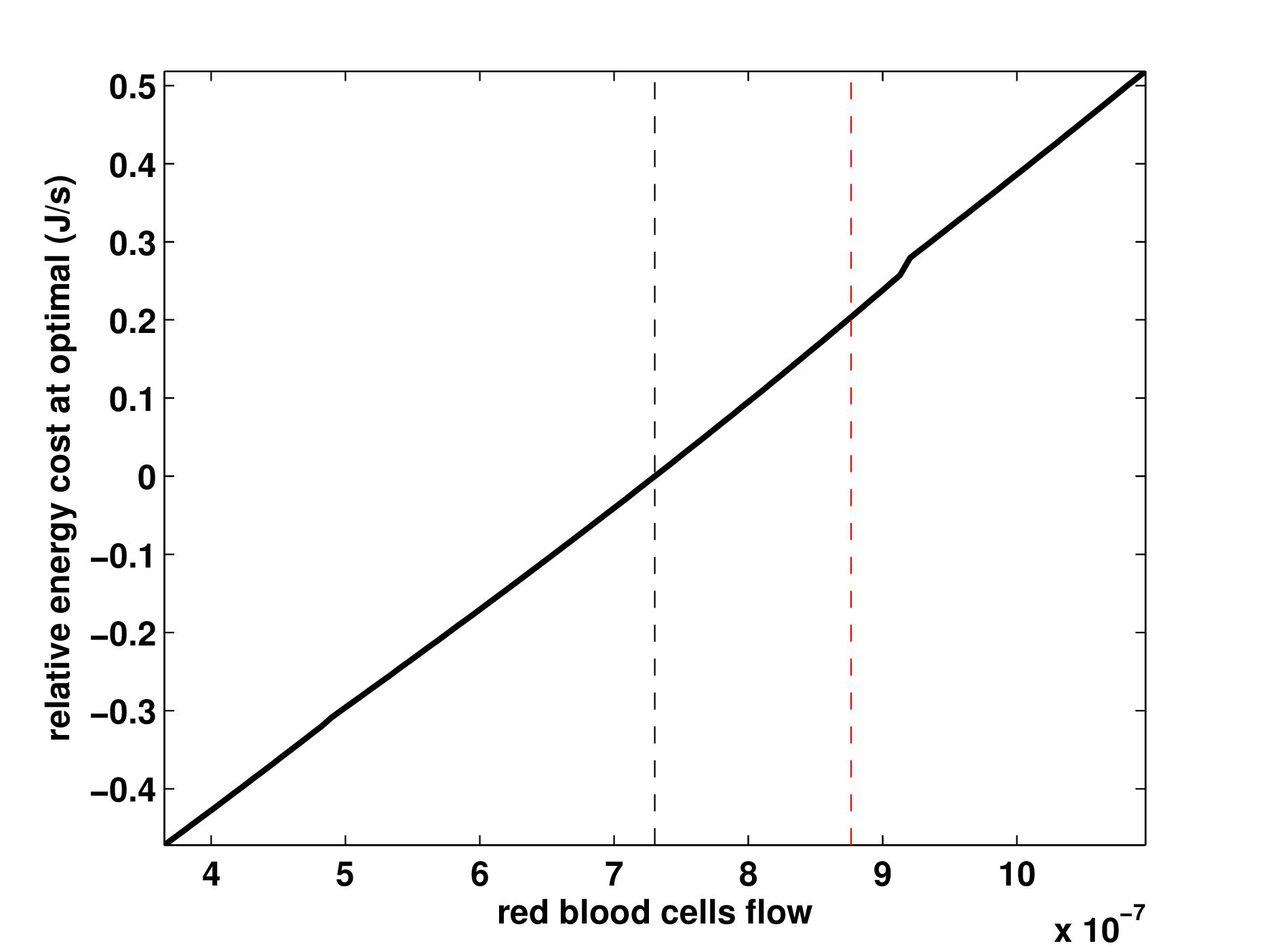}
D
\includegraphics[height=5cm]{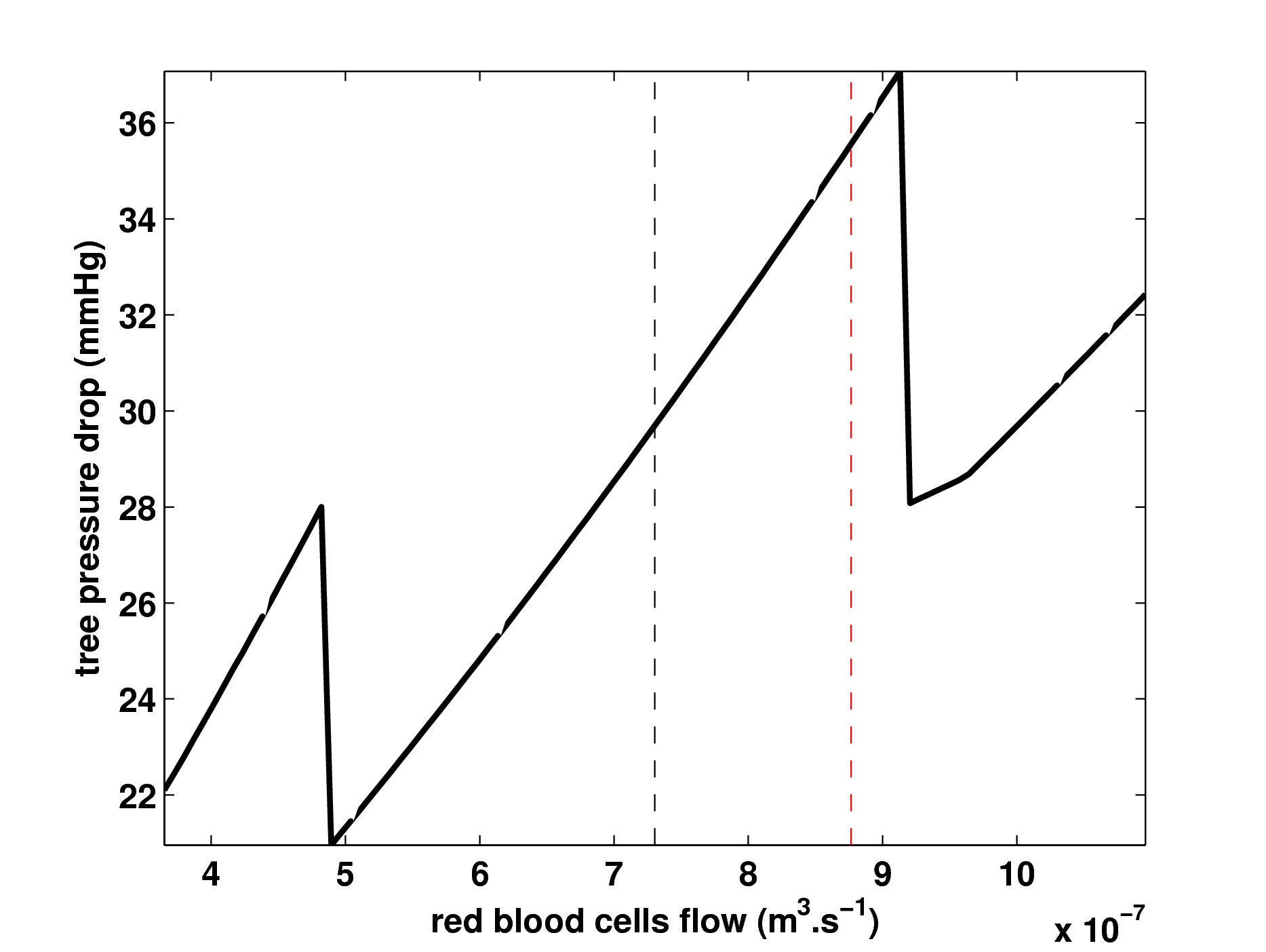}
\caption{Influence of red blood cells flow $\qrbc$ on optimal configuration relatively to the reference configuration $\qrbc = 7.3 \ 10^{-7} \ m^3.s^{-1}$ (black dashed line). The red dashed line represents a twenty percents increase of the red blood cells flow, mimicking lower hemoglobin saturation at an altitude of $3800 \ m$. The jumps correspond to a change in the total number of generations. A: Optimal hematocrit versus red blood cells flow; B: optimal homothety ratio versus red blood cells flow; C: relative energy cost versus red blood cells flow. D: pressure drop throughout the tree in the optimal configuration.}
\label{HhvsRBCflow}
\end{figure}

\noindent {\bf Red blood cells flow constraint.}
The red blood cells flow constraint is dependent on the organ oxygen needs or on the carrying capacity of red blood cells. Oxygen needs depend on the organ (muscles, brain), the individual (normal, athlete), the species, the respiration regime (exercise, rest), etc. Difference in oxygen carrying capacity of red blood cells can be a consequence of variability between individuals or between species, of different environments (sea-level or altitude), of pathologies (hypoxia), etc. The influence of the red blood cells contraint is plotted on figure \ref{HhvsRBCflow}, red blood cells flow constraint ranges from $50 \ \%$ less to $50 \ \%$ more than the reference value $\qrbc = 7.3 \ 10^{-7} \ m^3.s^{-1}$ \cite{comolet}.

Optimal hematocrit remains physiological through all the range of red blood cells flow tested. Optimal geometry remains $h=0.79$ on a wide range of red blood cells flow constraints. In that range the optimal hematocrit is increasing with red blood cells flow constraint, going from $0.34$ to $0.48$. Outside of this range, the optimal geometry changes, with a smaller $h$ of about $0.775$ for low red blood cells constraints and larger $h$ of about $0.80$ for high red blood cells flow constraints. The change in optimal geometry goes along with a change in the number of generations and induces jumps in the optimal hematocrit, see figure \ref{HhvsRBCflow}A and \ref{HhvsRBCflow}B. 

More red blood cells transport implies a higher energy cost, almost proportional to the red blood cells flow constraints, as shown on figure \ref{HhvsRBCflow}C. The supplementary cost is due to two effects: a higher blood viscosity (higher hematocrit) and a higher blood velocity. The pressure drop needed throughout the tree to sustain the flow of red blood cells is more sensitive to the increase of hematocrit than to the increase of blood velocity, see figure \ref{HhvsRBCflow}D. Interestingly, on the contrary of the energy cost and pressure drop, that are proportional to the length of the tree, the optimal hematocrit and homothety ratio do not depend on that length.

The red dashed lines on figures \ref{HhvsRBCflow} represent an increase of red blood cells flow of about $20 \ \%$ that would compensate the decrease of oxygen partial pressure in an altitude of about $3800 \ m$ (Andean or Tibetan plateaus). In that new environment, optimal hematocrit is about $0.48$, thus about $12 \ \%$ larger than sea-level hematocrit; the $8 \ \%$ missing are gained by blood flow increase. The optimal geometry is the same as that of sea-level. The physiological response to altitude acclimatization is similar to the response to hypoxia and stimulate erythropoiesis, hence increasing hematocrit value \cite{villafuerte,zubieta}. At first sight, our results indicate that this response may be advantageous, since it brings the hematocrit closer to the optimal predicted and allows to keep sufficient oxygen flow. However, that new optimal goes with an additional energy cost of about $20 \ \%$, due both to highest blood viscosity and speed. This supplementary energy cost could overburden heart and metabolism in general and may play a role to the development of chronic mountain thickness \cite{reeves}. Indeed, acclimatization response is maladaptive and counter-selected in Tibetan populations that have colonized high altitude plateaus for more than $25 000$ years \cite{storz, storz2}. Evolutive response to altitude tends to bring back to a sea-level hematocrit and blood flow, and to optimize parts of the oxygen pathway not accounted for in our model, such as the oxygen/hemoglobin binding properties or the mitochondrial distribution \cite{storz}, thus canceling the supplementary cost for oxygen transport in the network.\\

\noindent{\bf Allometric laws.}
Our model and data corresponds to a subtree of mammal arterial network. Indeed, apart for a few exceptions, such as physiological change in red blood cells concentration -released for example by spleen in horses during exercise \cite{fedde}- or specifically adapted red blood cells geometry and behavior -camels and water control \cite{weibel}-, very slight differences between red blood cells sizes, behavior and concentration exist between mammals species \cite{weibel}. Geometries of capillaries are also very similar \cite{west1}. Consequently, blood properties and flow through the subtree we studied can reasonably be considered independent on mammal species.
For two mammals species with different sizes, men or rabbit for example, the difference in arterial network structure occurs on the number of subtrees and on total blood flow: a larger mammal will have more subtrees than a smaller one, and its total blood flow will be correspondingly larger according to a scaling law \cite{west1}. Nevertheless, thanks to the tree-like structure of the arterial network, the volumetric proportion of those subtrees in the whole arterial network depends only slightly on the size of the network. Thus the influence of that subtree on the selection of the optimal hematocrit in the arterial network is similar whatever the size of the mammal, at least if they are big enough for their blood network to have $5 \ mm$ diameter vessels. For smaller mammals (mouse), our results would also hold: in the optimal configuration with geometry $h=0.79$, the contribution of each generation on the optimal hematocrit is the same whatever the generation, see figure \ref{optHBr}A. Consequently, optimal configuration is not dependent on the root radius $r_0$. This indicates that most of our results and conclusions apply to mammals species of any size.

\subsection{Model hypotheses and data}

As shown by the following example based on Poiseuille's law, constraint choice is important: we consider a vessel whose hydrodynamic resistance is $R$. If the flow $Q$ is constrained, then the minimization of the dissipation $P = R Q^2$ leads to minimize the hydrodynamic resistance $R$ of the vessel. If the pressure drop $\bigtriangleup p = R Q$ is constrained, then the minimization of the dissipation $P = R Q^2 = (\bigtriangleup p)^2 / R$ leads to maximizing $R$. The optimal vessel has an infinite diameter for the first case, and a zero diameter for the second case. While some constraints could be easily turned out because the resulting optimal configurations are non physiological, some others could lead to optimal configurations seemingly physiological but nonetheless faulty. Consequently, the evolutive constraint should be carefully chosen from established evolutionary theory: Fisher fitness theory \cite{fisher}, its many extensions \cite{lack, parker, vercken} and symmorphosis theory \cite{weibelsym} which links physiology and evolution. We study here the selection of the vascular network relatively to its function as oxygen supplier, however blood network has many other functions that are not accounted for in this work. Thus, the predictions of any model need to be interpreted carefully in the frame of the hypotheses used to build that model.

We assumed organ oxygen needs to be constant and we search for the blood network that is best adapted to fulfill those needs. The most realistic scenario would be that organs and blood network have evolved together, oxygen needs going along with oxygen supplies. Our model does not follow this evolutive dynamics. However, evolution has probably selected individuals whose amount of oxygen supply at a time roughly fitted the amount of oxygen needed at that time. Consequently, the optimal configuration reached today by natural selection is likely to coincide with the optimal configuration found by our model, in the limit of our hypotheses.

To get better predictions, the model can be improved. Typically, smooth vessels probably reduce the dissipation and influence pressure drops. Similarly, we do not take into account how bifurcations, asymmetric branching, blood inertia, turbulence or pulsations alter blood flow. However, the model developed in this paper accounts for the most important features that drives the cost and function of blood arterial network: the network tree structure and the rheology of blood. 
Another simplification appears by the use of Pries semi-empirical law that corresponds to high sheared flow. That law is thus more adapted to geometry implying high shear rates (i.e. for $h<1/2^{1/3}$). However, the role of the depleted layer on optimal hematocrits and geometries remain small.
Nevertheless, the predictions of our model are close to the physiological data and suggest that we accounted for the principal forces that drove the evolutive dynamics of blood arterial network, at least for its proximal parts. Yet, the optimal geometry of the highly dissipative microcirculation and its role on the optimal hematocrit were not taken into account in our model. Actually, in the microcirculation, blood cannot be modeled anymore as a continuous medium because the red blood cells size and the vessels diameters are of the same order of magnitude. Consequently, blood has to be modeled as a fluid interacting with deformable bodies using high-end numerical techniques \cite{ai, mauroyRBC, moreau, richardson2, veerap1, veerap2}. 

We optimized relatively to a couple of variables: a variable linked to the geometry, the homothety ratio $h$, and a variable linked to the fluid, the hematocrit $H$. We assumed that both variables were constant through the whole network, however these quantities are not of the same kind: (discharge) hematocrit is a global characteristic, because blood fills and flows in the whole network; the homothety ratio, however, refers both to a local characteristic of each bifurcation and to a global characteristic of blood, i.e. the volume of blood and the cost of its maintenance. This brings in the possibility of different regimes for the parameter $h$ depending on the location or size of the vessels considered. 

We estimated the costs for the metabolism to maintain and to renew red blood cells with data from the literature, sometimes using rough computations. To reach better estimations, dedicated experiments should be made. Moreover, we measured the costs as if the glucose and oxygen consumed by the red blood cells, mostly by anaerobic glycolysis, would have been consumed by the metabolism by aerobic glycolysis, more efficient than anaerobic pathway. This hypothesis is based on the fact that most of the glycolysis in mammals is aerobic \cite{weibel}. This can however be refined by taking into account the proportion of both pathways, depending on the respiration regime for example.

\section{Conclusion}

In this work, we developed a theoretical and numerical model of a subtree of the arterial network of mammals (diameters from $5 \ mm$ down to $ 50 \ \mu m$) and we optimized two major features of the network: its geometry and the red blood cells concentration in blood (hematocrit). 
We propose that the arterial network evolved while being constrained by its function as an organ, thus following the principles introduced by symmorphosis. To illustrate this hypothesis, we focused our study on one of the main function of blood network: oxygen supply to the organs. We considered an idealized organ with a given oxygen need and we optimized blood network geometry and hematocrit with the constraint that it must fulfill the organ oxygen need. We integrated in our model the non-Newtonian behavior of blood and its maintenance cost. Our predictions are compatible with the physiology: we found an optimal hematocrit of  $0.43$ and an optimal homothety ratio of about $0.79$ for an average human. These results indicate that the role of the intermediate range of arterial network generations on the selection of an optimal hematocrit may be strong. Finally, our results suggest that pressure drops in the arterial network should be regulated in order for oxygen supply to remain optimal.

Our model is also able to compute the optimal hematocrit in a single vessel. We show that the vessel optimal hematocrit is an increasing function of the mean shear rate in that vessel, with a steep transition for shear rates  between $10 \ s^{-1}$ and $100 \ s^{-1}$ from a low optimal hematocrit to a high optimal hematocrit. This distribution is slightly altered if \fah effect is taken into account. We also showed that the mean shear rate in the different levels of a fractal tree network follows a scaling law that drives the behavior of the mean shear rate along the generations of the tree. Consequently this scaling law also drives the variation of the vessels optimal hematocrits along the tree generations. The global optimal hematocrit in the tree network is a compromise between the influences of each vessels on the global cost, and stands between the vessel optimal hematocrit in the tree root and the vessel optimal hematocrit in the tree leaves.

\section*{Acknowledgments}
The authors would like to thank Philippe Dantan, Patrice Flaud and Daniel Qu\'emada for very interesting and helpful discussions. The authors would humbly like to dedicate this paper to the memory of Daniel Qu\'emada.

\end{document}